# Blockchain-Driven Solutions for Carbon Credit Trading: A Decentralized Platform for SMEs

Yun-Cheng Tsai

**Abstract**

The increasing demand for sustainability and compliance with global carbon regulations has posed significant challenges for small and medium-sized enterprises (SMEs). This paper proposes a blockchain-based decentralized carbon credit trading platform tailored for SMEs in Taiwan, aiming to simplify the complex carbon trading process and lower market entry barriers. Drawing upon the Diffusion of Innovations theory and transaction cost economics, we illustrate how blockchain technology can reduce informational asymmetry and intermediary costs in carbon markets. By integrating Ethereum-based smart contracts, the platform automates transactions, enhances transparency, and reduces administrative burdens—addressing key obstacles such as technical complexity and market risks. A controlled experimental design was conducted to compare the proposed system with a conventional centralized carbon trading platform. Statistical analysis confirms its effectiveness in minimizing time and expenses while ensuring compliance with the Carbon Border Adjustment Mechanism (CBAM) and the Clean Competition Act (CCA). User satisfaction was measured using the Kano model, with the results identifying essential features and prioritizing future enhancements. This study contributes a more comprehensive solution for SMEs seeking to achieve carbon neutrality, underscoring the transformative potential of blockchain technology in global carbon markets.

---

## 1. Introduction

### 1.1 Background

Global efforts to combat climate change have accelerated the development of carbon trading markets as a pivotal mechanism to reduce greenhouse gas (GHG) emissions. Regulatory frameworks such as the European Union's Carbon Border Adjustment Mechanism (CBAM) [5] and the United States' Clean Competition Act (CCA) [6] underscore the importance of efficient and transparent carbon credit systems. These regulations aim to minimize "carbon leakage" by ensuring that imported goods are subject to similar carbon costs as those produced domestically, thereby promoting fairness and maintaining competitive balance [1], [5]. The Intergovernmental Panel on Climate Change (IPCC) further emphasizes that timely and ambitious CO2 reduction measures are crucial to limiting global warming [7].

However, small and medium-sized enterprises (SMEs) often encounter significant barriers when attempting to participate in carbon markets. These obstacles include:

1. **High Technical Thresholds**: Many SMEs lack the expertise or resources to manage blockchain or carbon credit portfolios effectively [4].

2. **Market Risks**: Volatile carbon pricing and limited market access heighten the financial risk for SMEs [1].

3. **Administrative Complexities**: Regulatory compliance, involving meticulous documentation and auditing, can overwhelm companies with limited operational capacity [5].

As illustrated in **Fig. 1**, the CBAM is scheduled to begin its trial implementation on October 1, 2023, requiring importers to provide product carbon emissions data. After the transition period, the payment system will be officially launched in 2026, initially targeting five high-carbon-emitting industries. By 2034, the EU plans to completely abolish ETS free quotas and fully incorporate CBAM into other sectors 555.

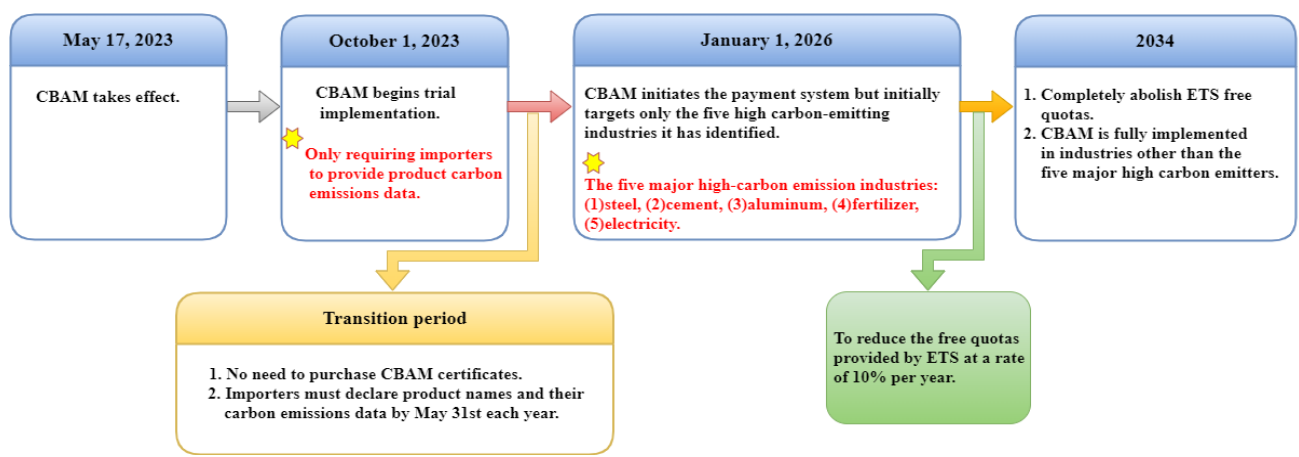

Fig. 1: Implementation timeline of the Carbon Border Adjustment Mechanism (CBAM) [5].

## 1.2 Literature Review

Studies have highlighted that blockchain's decentralized, tamper-proof properties can mitigate information asymmetry in various markets, including carbon trading [10], [11]. Prior research on blockchain-based carbon trading platforms indicates that smart contracts can automate transactions and reduce reliance on intermediaries [4], yet practical implementations often face scalability, regulatory, and governance challenges [12]. Comparisons among different blockchain frameworks (e.g., Hyperledger, Corda, Ethereum) suggest that Ethereum's mature ecosystem and developer support facilitate faster prototyping [3], though Hyperledger has been explored for enterprise-grade permissioned scenarios [11]. Additionally, from a theoretical standpoint, **transaction cost economics** posits that lowering coordination costs encourages broader market participation [13], while **Diffusion of Innovations** theory explains how perceived complexity or relative advantage affects the rate of new technology adoption [14]. Building on these insights, our research aims to design a platform that lowers the entry barriers for SMEs through a user-friendly interface and regulatory compliance automation.

As shown in **Fig. 2**, the Clean Competition Act was introduced by the U.S. Senate on June 7, 2022, and is scheduled for implementation in 2024. The legislation imposes carbon tariffs on high-carbon industries (e.g., petrochemicals, cement, steel, aluminum, fertilizers), alongside a detailed carbon tax calculation method. These regulations are designed to gradually reduce the baseline for permissible emissions, increasing carbon taxes over time if emissions exceed the established threshold 666.

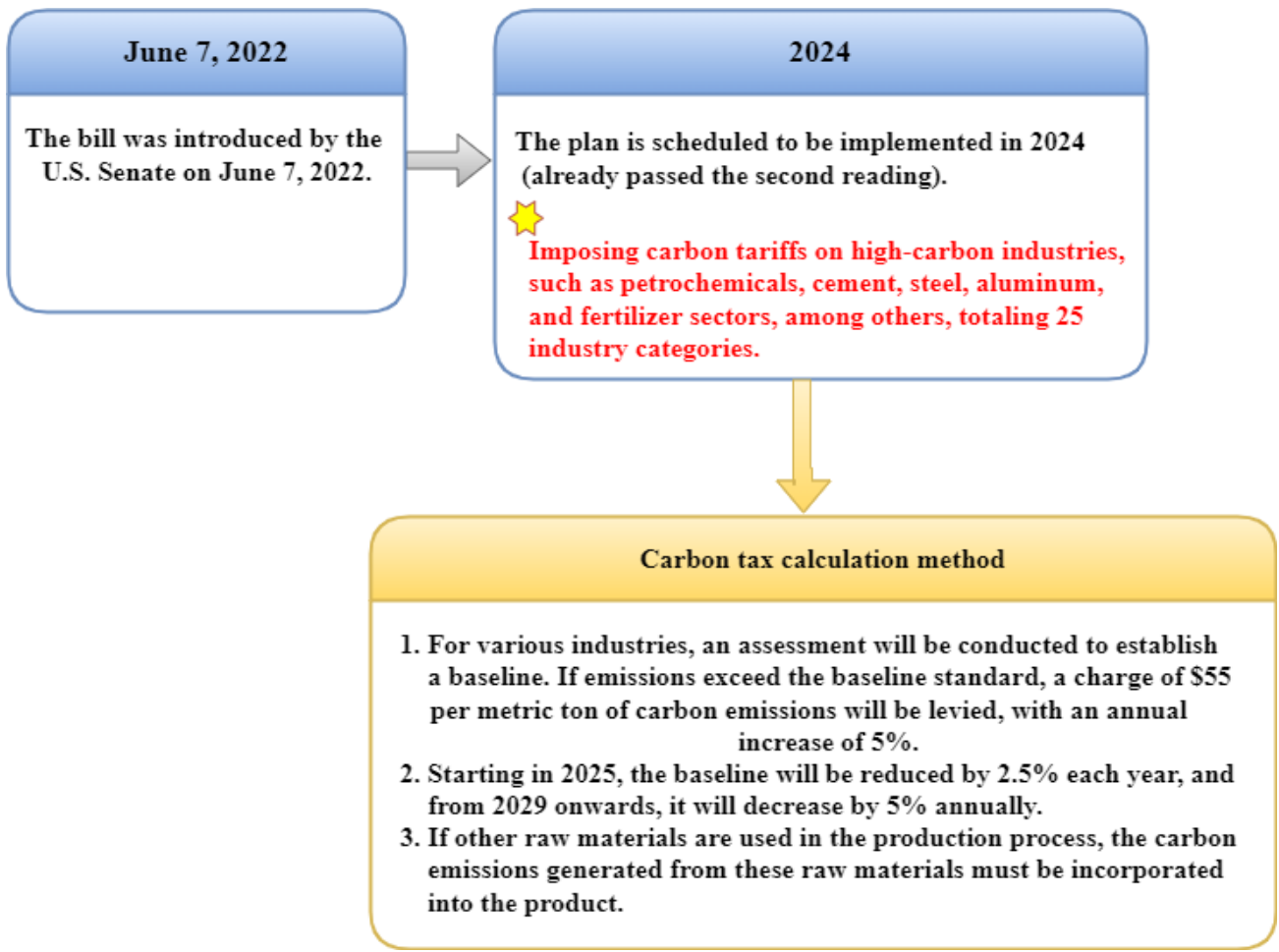

Fig. 2: Timeline and carbon tax calculation method under the Clean Competition Act [6].

**1.3 Research Motivations and Objectives**

Leveraging blockchain's potential [8] and informed by relevant theoretical frameworks, this research aims to:

1. **Develop a Decentralized Platform**: Provide SMEs in Taiwan with a robust, user-friendly blockchain-based system for carbon credit trading.

2. **Enhance Transparency and Security**: Utilize the Ethereum blockchain to record all transactions immutably, thereby building trust among participants [2].

3. **Ensure Regulatory Compliance**: Align platform functionalities with CBAM [5], CCA [6], and potentially the Carbon Offsetting and Reduction Scheme for International Aviation (CORSIA) [8].

4. **Validate Performance and User Satisfaction**: Evaluate the platform's efficiency, cost-effectiveness, and user acceptance through a controlled experimental setup, accompanied by Kano model analysis [9].

By addressing these objectives, our study provides both theoretical and practical contributions, highlighting how blockchain can reduce transaction costs and facilitate broader technology adoption among SMEs [13], [14].

## 2. Materials and Methods

### 2.1 Research Context

The platform prototype targets SMEs in Taiwan's manufacturing and service sectors, which often operate under tight profit margins [5], [7]. A decentralized system reduces reliance on costly intermediaries while offering a standardized method for carbon credit transactions. Given the international scope of carbon regulations, our design also accounts for potential future integration with frameworks such as CORSIA [8]. By embedding modular compliance checks within smart contracts, the platform can adapt to regulatory updates over time.

Fig. 3 illustrates the organizational structure and responsibility flow among the key stakeholders, including Rajasthan Renewable Energy Corporation (RREC), Balkishan Industries Limited (BIL), Suzlon Energy Limited (SEL), and other relevant entities. As shown, RREC oversees document verification and project approvals, while SEL monitors daily power generation, and BIL manages electricity distribution to its manufacturing unit.

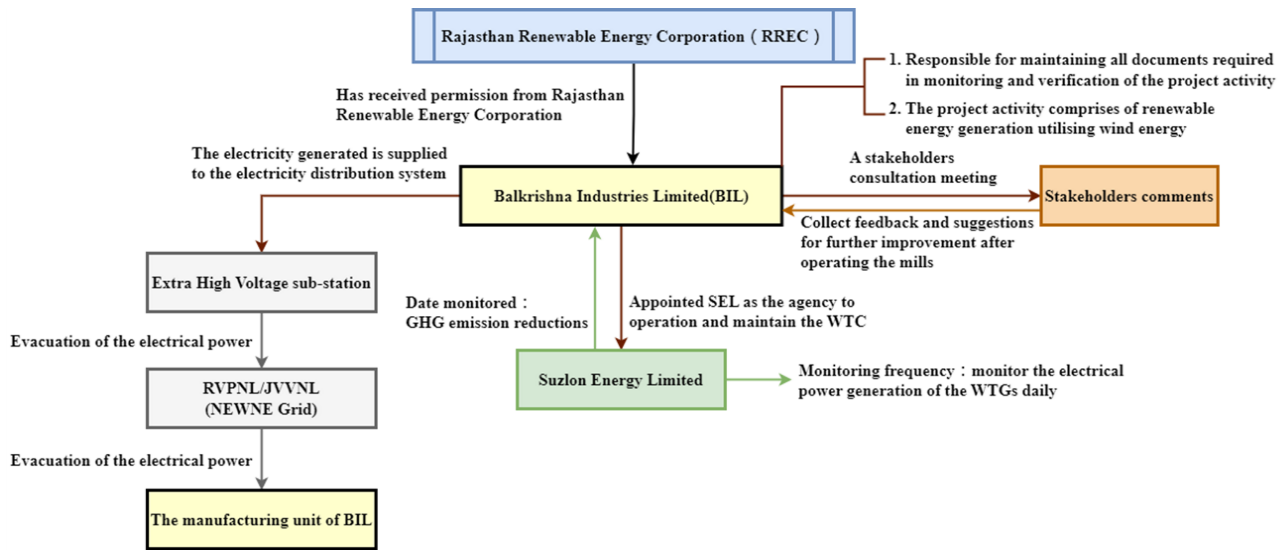

Fig. 3: Process flow chart illustrating stakeholder roles, documentation procedures, and monitoring responsibilities in the renewable wind energy project.

### 2.2 Theoretical Underpinnings

1. **Diffusion of Innovations (DOI)**: Guides our understanding of how perceived complexity, relative advantage, and compatibility influence SME adoption of blockchain-based solutions [14].

2. **Transaction Cost Economics (TCE)**: Explains how reducing intermediary and auditing costs can incentivize market participation and promote efficiency in carbon trading [13].

These theories form the conceptual basis for hypothesizing that a user-friendly blockchain platform with automated compliance checks can significantly lower barriers to entry and operational costs for SMEs.

### 2.3 System Architecture

The platform's architecture is built upon the Ethereum blockchain, known for its robust smart contract capabilities and sizable developer community [2], [3]. Major components include:

1. **Smart Contracts**: Developed in Solidity, governing core functionalities such as credit issuance, trading, transfers, and cancellations [4].

2. **User Interface**: A React.js web application provides an intuitive interface for non-technical users, including wallet integration and real-time transaction tracking [2].

3. **Blockchain Network**: Either Ethereum Mainnet or a private instance, depending on transaction volume, security requirements, and cost constraints [3].

4. **MetaMask Integration**: Users authenticate and sign transactions via MetaMask for secure interactions with the blockchain.

5. **Off-Chain Database**: Certain user profiles or documents are stored off-chain to reduce gas costs, while essential transaction data remains immutable on-chain [8].

A simplified illustration (Fig. 1) shows how the React.js front-end communicates with Solidity smart contracts on the Ethereum network, eliminating the need for a centralized clearinghouse [12].

### 2.4 Smart Contract Implementation

To demonstrate automated carbon credit trading feasibility, we implemented four core Solidity contracts [3], [4], [5]:

- **CarbonCreditToken.sol**: An ERC-721 compliant token representing carbon credits, each symbolizing 1 metric ton of $CO_2e$.

- **MarketPlace.sol**: Handles listing and purchase of carbon credits. Sellers list tokens at desired prices; buyers purchase them directly, triggering a token transfer.

- **CancellationAndRefund.sol**: Manages refund policies for unutilized or invalidated credits, ensuring fairness and regulatory alignment.

- **RewardMechanism.sol**: Encourages early participation via loyalty incentives (e.g., bonus credits or platform tokens).

All transactions (buyer/seller addresses, credit amounts, timestamps) are recorded on-chain, regulated by role-based permissions to mitigate fraud [1]. Furthermore, a third-party auditor can verify token authenticity and transaction logs to ensure compliance with local and international standards [5], [6].

### 2.5 Experimental Design

#### 2.5.1 Experimental Groups

We employed a **controlled experimental design** with two groups:

- **Treatment Group (Blockchain Platform)**: 15 SMEs used the proposed decentralized platform for simulated carbon credit trading.

- **Control Group (Conventional System)**: Another 15 SMEs (similar industry composition) used a centralized carbon trading platform with manual verification and broker-mediated transactions.

By comparing these two groups, we isolate the effect of blockchain-based automation versus traditional market structures.

**2.5.2 Test Environment**

A private Ethereum test network (using Ganache) was used to evaluate functionality and performance [3]. This environment supports rapid iteration and cost-free transactions during development. The control group performed transactions via a conventional broker or exchange interface, simulating real-world processes where requests and settlements are typically handled by an intermediary.

**2.5.3 Data Collection and Metrics**

1. **Transaction Time (seconds)**: Duration from user-initiated "Purchase" to on-chain confirmation or centralized broker confirmation [3], [4].

2. **Transaction Cost (USD)**: Calculated via Ethereum gas fees plus any platform fee; for the control group, broker or exchange fees were aggregated [2], [15].

3. **System Usability (Kano Model)**: Evaluated using structured surveys; features are classified into Must-Be, One-Dimensional, and Attractive categories [9].

4. **Compliance Effectiveness**: Verified alignment with CBAM [5], CCA [6], and future applicability to CORSIA [8] through auditor review of transaction records.

5. **Statistical Significance**: t-tests ($\alpha = 0.05$) were conducted to determine if the differences in transaction times and costs between groups were statistically significant.

**2.5.4 Sample Selection and Limitations**

Each group comprised 15 SMEs drawn from manufacturing, food services, and textiles. While this sample size offers preliminary insights, the limited scope may affect generalizability. Future research could expand sample size or focus on a single industry for deeper analysis.

---

**3. Results and Discussion**

**3.1 Platform Performance**

**3.1.1 Transaction Efficiency**

Statistical analysis (independent samples t-test) revealed that the **treatment group** achieved a significantly lower average transaction time (M = 38.2s, SD = 5.0s) compared to the **control group** (M = 88.7s, SD = 12.3s), with $p < 0.01$. This ~57% reduction can be attributed to automated smart contracts that eliminate manual verifications and intermediary processing [4].

| Metric | Control Group (Mean ± SD) | Treatment Group (Mean ± SD) |
|---|---|---|
| Transaction Time (s) | 88.7 ± 12.3 | 38.2 ± 5.0 |
| Transaction Cost (USD) | 3.50 ± 1.10 | 1.90 ± 0.65 |

**3.1.2 Transaction Cost**

Gas fees on the private Ethereum network were nominal, averaging $1.90 per transaction, whereas the control group incurred broker/exchange fees averaging $3.50, leading to a cost reduction of approximately 46%. These differences are statistically significant at $p < 0.05$, supporting the hypothesis that blockchain automation lowers direct financial overhead [13].

### 3.1.3 Scalability Considerations

If transaction volume scales dramatically, Ethereum Mainnet's network congestion and gas fee volatility could hinder performance [3]. Layer-2 solutions or alternative blockchains (e.g., Polygon, Solana) might mitigate such bottlenecks [8]. Additionally, implementing an interoperable framework could enable cross-chain asset transfers, improving market liquidity [12].

## 3.2 User Satisfaction

Based on the Kano model [9], surveys indicated:

- **Must-Be Features**: Secure login, transparent transaction logs, compliance reporting.
- **One-Dimensional Features**: Fast processing time, low fees, real-time credit price tracking.
- **Attractive Features**: Mobile app integration, reward incentives, and advanced analytics.

Over 80% of the treatment group users rated the interface as "intuitive" or "very intuitive," aligning with **Diffusion of Innovations** aspects where perceived ease-of-use fosters technology adoption [14]. This user-centric design, combined with real-time feedback on transactions, likely contributed to higher satisfaction levels compared to the control group.

## 3.3 Environmental and Economic Impacts

Lower entry barriers encourage SMEs to engage in carbon offset activities, thereby contributing to broader GHG reduction goals [1], [5], [7]. The platform:

1. **Reduces Transaction Costs**: Minimal brokerage fees and automated record-keeping allow SMEs to allocate more resources to actual carbon abatement [5].
2. **Broadens Market Access**: Simplified processes facilitate participation in global carbon markets, potentially fostering a more liquid and transparent trading ecosystem [6].
3. **Ensures Regulatory Assurance**: Smart contracts can embed compliance checks for CBAM and CCA requirements, diminishing the risk of non-compliance [5], [6].

## 3.4 Challenges and Limitations

1. **Regulatory Coordination**: Different international regulations may require ongoing updates to the smart contract logic [5], [6].
2. **Blockchain Adoption**: Many SMEs lack familiarity with wallets, gas fees, or private keys, requiring training and user-friendly design [2], [3].
3. **Governance and Auditing**: Continuous oversight by third parties is essential to validate carbon credit authenticity, necessitating robust governance mechanisms [5], [12].
4. **Initial Setup and TCO**: Despite lower per-transaction costs, organizations must consider infrastructure, maintenance, and staff training. A more detailed total cost of ownership (TCO) analysis is recommended [15].

## 3.5 Comparison with Traditional Systems

Results underscore significant improvements in accessibility, transparency, and efficiency [1], [5]. However, conventional systems often benefit from well-established infrastructures and deeper market liquidity, an aspect that the proposed platform must address through strategic partnerships and potential cross-chain expansions [8].

---

## 4. Conclusion

This research presents a blockchain-based carbon credit trading platform specifically designed for SMEs, rooted in the Diffusion of Innovations theory and transaction cost economics. By leveraging Ethereum smart contracts, the system automates carbon credit transactions, reduces operational costs, and provides transparent records to bolster trust among market participants [2], [3], [4]. Controlled experimentation indicates a statistically significant reduction in both transaction time and fees. Kano model analysis further demonstrates high user satisfaction [9], underscoring the platform's user-centric approach.

### 4.1 Contributions and Future Work

Key contributions include:

1. **Theory-Driven Design**: Integrating Diffusion of Innovations and TCE to address adoption and cost-related barriers [13], [14].

2. **Robust Experimental Comparison**: Implementing a controlled design to compare blockchain vs. conventional trading platforms, offering quantitative evidence.

3. **Regulatory Compliance Framework**: Embedding CBAM, CCA, and CORSIA protocols directly in smart contracts for automated reporting [5], [6], [8].

Moving forward:

- **KYC/AML Integration**: Strengthening identity verification to meet global anti-money laundering requirements.

- **Advanced Analytics**: Incorporating predictive pricing models and portfolio recommendations [7].

- **Scalability & Interoperability**: Investigating Layer-2 or cross-chain solutions for improved performance in high-volume scenarios [8], [12].

- **Governance Models**: Exploring decentralized autonomous organization (DAO) mechanisms to manage protocol upgrades and dispute resolution.

By offering a feasible path for SMEs to engage in carbon trading, the proposed platform promotes broader sustainability goals and supports the global shift toward carbon neutrality [7]. Our findings illustrate the potential of blockchain to reshape carbon markets, while acknowledging that regulatory, technological, and governance challenges warrant ongoing research.

---

**Acknowledgments**

---